\documentclass[prd,showpacs,preprintnumbers,amsmath,amssymb]{revtex4}

\usepackage{graphicx}
\usepackage{dcolumn}
\usepackage{bm}

\draft

\begin{document}

\title{Schwinger Pair Production at Finite Temperature in QED}
\author{Sang Pyo Kim}\email{sangkim@kunsan.ac.kr}
\affiliation{Department of Physics, Kunsan National University,
Kunsan 573-701, Korea} \affiliation{Asia
Pacific Center for Theoretical Physics, Pohang 790-784, Korea}

\author{Hyun Kyu Lee}\email{hyunkyu@hanyang.ac.kr}
\author{Yongsung Yoon}\email{cem@hanyang.ac.kr}
\affiliation{Department of Physics, Hanyang University, Seoul
133-791, Korea}

\date{\today}

\begin{abstract}
We use the evolution operator method to find the Schwinger
pair-production rate at finite temperature in scalar and spinor
QED by counting the vacuum production, the induced production and
the stimulated annihilation from the initial ensemble. It is shown
that the pair-production rate for each state is factorized into
the mean number at zero temperature and the initial thermal
distribution for bosons and fermions.
\end{abstract}
 \maketitle

\section{Introduction}

Vacuum polarization and pair production have been issues of continuous concern since the early works by Sauter, Heisenberg
and Euler, and Weisskopf \cite{Sauter}, and then by Schwinger \cite{Schwinger} (for a review and references, see Ref.
\cite{Dunne}). The task of directly computing, without relying on the electromagnetic duality, the effective action in
electric field backgrounds, however, has been a challenging problem due to the vacuum instability. Dunne and Hall used the
resolvent method to directly find the effective action in time-dependent electric fields \cite{Dunne-Hall}. In the previous
paper \cite{KLY}, employing the evolution operator method, we found the exact one-loop effective actions of scalar and
spinor QED at zero temperature in a constant or a pulsed electric field of Sauter-type, which satisfy the exact relation $2
{\rm Im} {\cal L}_{\rm eff} =\pm \sum_{n} \ln (1\pm \overline{\cal N}_{n})$ (with $+$ for scalar and $-$ for spinor)
between the imaginary part of the effective Lagrangian density ${\cal L}_{\rm eff}$ and the mean number of created pairs
$\overline{\cal N}_{n}$ at state $n$. Even finding the pair production rate by time-dependent or spatially localized
electric fields is methodologically nontrivial, which has recently been intensively studied \cite{Kim-Page,DSWG,KRX}.

To calculate the effective action and thereby Schwinger pair production at finite temperature is another challenging
problem in QED. In a constant pure magnetic field the QED effective action was studied at finite temperature
\cite{Dittrich} and  at finite temperature and density \cite{EPS}. However, the presence of an additional electric field
raised Schwinger pair production at debate depending on the formalism employed. Some calculations in thermal field theory
reported that the effective action in both a constant magnetic and electric field had an imaginary part having dependence
on temperature \cite{Loewe-Rojas,GKP,Hallin-Liljenberg}. However, in the real-time formalism, no imaginary part was found
in the QED effective action in the presence of both magnetic and electric field \cite{Elmfors-Skagerstam}. The recent
calculation of effective action in the imaginary-time formalism shows an imaginary part only at two-loop but not at
one-loop in both a constant electric and magnetic field \cite{Gies99}. In nonequilibrium quantum field theory of scalar
QED, a calculation in the real-time formalism shows thermal enhancement of pair production \cite{Kim-Lee07}.

In this paper, using the evolution operator method, we find the pair-production rate at finite temperature in
time-dependent electric fields both in scalar and spinor QED. The evolution operator, unitarily transforming the
particle and antiparticle operators from the ingoing vacuum to the outgoing vacuum, carries all the information of
quantum evolution. In fact, the evolution operator is completely determined by the Bogoliubov coefficients. The
advantage of the evolution operator is the readiness to calculate the probability for transitions among multiparticle
states.  This allows us to compute the mean number of created pairs at finite temperature in scalar and spinor QED by
counting the pairs from the vacuum and the induced production and the stimulated annihilation in a thermal ensemble of
bosons and fermions. We find that the mean number of created pairs is factorized into the mean number of created pairs
at zero temperature and the initial thermal distribution for bosons and fermions.

The organization of this paper is as follows. In Sec. II,
rewriting the Bogoliubov transformation as a unitary
transformation by the evolution operator, we find the mean number
of created pairs at zero temperature in scalar and spinor QED. In
Sec. III, we calculate the mean number of created pairs at finite
temperature and apply it to the Sauter-type electric field.

\section{Evolution Operator and Pair Production at $T = 0$}

We first consider scalar QED for spinless charged bosons under an external electric field with the gauge field $A_{\mu}$.
The electric field is assumed to be acting on for a finite period of time so that the ingoing and the outgoing vacua are
well-defined. Thus, at $t_{\rm in} = - \infty$, before the external electric field being turned on, the scalar field is
free and the Hamiltonian takes the usual form
\begin{eqnarray}
H_{\rm in}^{\rm sc} =\int \frac{d^3 {\bf k}}{(2 \pi)^3} \omega_{{\bf
k}, {\rm in}} N_{{\bf k},{\rm in}} =\int \frac{d^3 {\bf k}}{(2
\pi)^3} \omega_{{\bf k}, {\rm in}} (a_{{\bf k}, {\rm in}}^{\dagger}
a_{{\bf k}, {\rm in}} + b_{{\bf k}, {\rm in}}^{\dagger} b_{{\bf k},
{\rm in}}), \label{sc-in-ham}
\end{eqnarray}
where $\omega_{{\bf k}, {\rm in}}$ is the initial frequency at
momentum ${\bf k}$. Here, the gauge is chosen $A_{\mu} = 0$, so
that the ingoing vacuum $\vert 0; t_{\rm in} \rangle$ is the
Minkowski vacuum $\vert 0 \rangle_{\rm M}$, annihilated by $a_{\bf
k} (t_{\rm in})$ and $b_{\bf k} (t_{\rm in})$ for each momentum
${\bf k}$. Similarly, the outgoing vacuum at $t_{\rm out} =
\infty$ is defined by $a_{\bf k} (t_{\rm out})$ and $b_{\bf k}
(t_{\rm out})$. These operators are related through the Bogoliubov
transformations \cite{Kim-Lee07}
\begin{eqnarray}
a_{{\bf k}, {\rm out}} = \mu_{\bf k} a_{{\bf k}, {\rm in}} +
\nu^*_{\bf k} b^{\dagger}_{{\bf k}, {\rm in}}, ~~~ b_{{\bf k},
{\rm out}} = \mu_{\bf k} b_{{\bf k}, {\rm in}} + \nu^*_{\bf k}
a^{\dagger}_{{\bf k}, {\rm in}}, \label{sc-out-in}
\end{eqnarray}
where $ |\mu_{\bf k}|^2 - |\nu_{\bf k}|^2 = 1$.

To express the outgoing vacuum as multiparticle states of the
ingoing vacuum, we rewrite the Bogoliubov transformations
(\ref{sc-out-in}) as a unitary transformation \cite{KLY}
\begin{eqnarray}
a_{{\bf k}, {\rm out}} (A) = U_{\bf k} (A) a_{{\bf k}, {\rm in}}
(0) U^{\dagger}_{\bf k} (A), ~~~ b_{{\bf k}, {\rm out}} (A) =
U_{\bf k} (A) b_{{\bf k}, {\rm in}} (0) U^{\dagger}_{\bf k} (A).
\label{sc-U-tr}
\end{eqnarray}
Here, the evolution operator $U_{\bf k}$ is factorized into the overall phase factor and the two-mode squeeze operator as \cite{Caves-Schumaker1,Caves-Schumaker2}
\begin{eqnarray}
U_{\bf k} (A) = e^{ i \theta_{\bf k} (a^{\dagger}_{{\bf k},
{\rm in}}a_{{\bf k}, {\rm in}} + b^{\dagger}_{{\bf k}, {\rm in}}
b_{{\bf k}, {\rm in}} + 1 ) } e^{ \xi_{\bf k} a^{\dagger}_{{\bf k}, {\rm in}}
b^{\dagger}_{{\bf k}, {\rm in}}} e^{
\frac{\gamma_{\bf k}}{2} ( a^{\dagger}_{{\bf k}, {\rm in}}
a_{{\bf k}, {\rm in}} + b^{\dagger}_{{\bf k}, {\rm in}} b_{{\bf k},
{\rm in}}+1 ) } e^{ - \xi_{\bf k}^* a_{{\bf k},
{\rm in}} b_{{\bf k}, {\rm in}} },
\end{eqnarray}
where
\begin{eqnarray}
e^{2 i \theta_{\bf k}} = \frac{\mu^*_{\bf k}}{\mu_{\bf k}}, \quad \xi_{\bf k} = \frac{\nu^*_{\bf k}}{\mu_{\bf k}}, \quad \gamma_{\bf k}
= - 2 \ln (\mu_{\bf k} ).
\end{eqnarray}

From the charge neutrality of the vacuum, equal numbers of
particles and antiparticles are produced at zero temperature and
they carry the opposite momenta due to the momentum conservation.
The multiparticle state of $n$-pairs can be concisely denoted as
$\vert n_{\bf k}, t \rangle = \vert n_{\bf k};\bar{n}_{\bf k}; t
\rangle$. The probability for $n$-pairs with momentum ${\bf k}$ to
be created from the vacuum is
\begin{eqnarray}
P_n ({\bf k}) = | \langle n_{\bf k}, {\rm out} \vert 0, {\rm in}
\rangle |^2  = | \langle n_{\bf k}, {\rm in} \vert
U^{\dagger}_{\bf k} \vert 0, {\rm in} \rangle |^2 = e^{\gamma_{\bf
k}} |\xi_{\bf k}|^{2n}.
\end{eqnarray}
Note that $P_0 = e^{\gamma_{\bf k}}$ and $P_1 = e^{\gamma_{\bf k}}
|\xi_{\bf k}|^{2}$ so that $P_n = P_0 (P_1/P_0)^n$ and $\sum_{n =
0}^{\infty} P_n = 1$ for each ${\bf k}$. Thus, at zero
temperature, the mean number of pairs created from the vacuum for
each momentum per unit volume is
\begin{eqnarray}
\overline{\cal N}_{\bf k}^{\rm sc}(T=0) = \sum_{n = 0}^{\infty} n
P_n ({\bf k}) = |\nu_{\bf k}|^{2}. \label{sc-mean}
\end{eqnarray}

Next, in spinor QED, before the interaction of an external
electric field, the spinor field is free without the gauge
potential $(A_{\mu}=0)$, and has the Hamiltonian given by
\begin{eqnarray}
H_{\rm in}^{\rm sp} =\sum_{\sigma} \int \frac{d^3 {\bf k}}{(2
\pi)^3} \omega_{n,{\rm in}}N_{n,{\rm in}} = \sum_{\sigma} \int
\frac{d^3 {\bf k}}{(2 \pi)^3} \omega_{n,{\rm in}} (b_{n,{\rm
in}}^{\dagger} b_{n,{\rm in}} + d_{n,{\rm in}}^{\dagger} d_{n,{\rm
in}}), \label{sp-in-ham}
\end{eqnarray}
where $b_{n,{\rm in}}$ and $d_{n,{\rm in}}$ are particle and antiparticle operators in the ingoing vacuum.  After the interaction of the electric field, the ingoing vacuum evolves to the outgoing vacuum, whose particle and antiparticle operator are $b_{n,{\rm out}}$ and $d_{n,{\rm out}}$.
The Bogoliubov transformations between the
ingoing and the outgoing particle and antiparticle operators,
$b_{n}, d_{n}$, are similarly given by
\begin{eqnarray}
b_{n,{\rm out}} = \mu_{n} b_{n,{\rm in}} + i\nu_{n}^{*}d_{n,{\rm
in}}^{\dagger}, ~~~ d_{n,{\rm out}} = \mu_{n} d_{n,{\rm in}} -
i\nu_{n}^{*}b_{n,{\rm in}}^{\dagger}, \label{sp-out-in}
\end{eqnarray}
where $|\mu_{n}|^{2}+|\nu_{n}|^{2}=1$ and $n=({\bf k},\sigma)$
with $\sigma = \pm 1/2$.
The Bogoliubov transformation can be also written as a unitary
transformation \cite{Fan}
\begin{eqnarray}
b_{n,{\rm out}} = U_{n}b_{n,{\rm in}}U_{n}^{\dagger}, ~~~
d_{n,{\rm out}} = U_{n}d_{n,{\rm in}}U_{n}^{\dagger},
\label{sp-U-tr}
\end{eqnarray}
where
\begin{equation}
U_{n}=e^{\xi_{n}b_{n,{\rm in}}^{\dagger}d_{n,{\rm in}}^{\dagger}}
e^{(\frac{\gamma_{n}}{2}+i\theta_{n})(b_{n,{\rm
in}}^{\dagger}b_{n,{\rm in}}+d_{n,{\rm in}}^{\dagger}d_{n,{\rm
in}}-1)} e^{e^{2i\theta_{n}}\xi_{n}^{*}b_{n,{\rm in}}d_{n,{\rm
in}}}. \label{sp-U}
\end{equation}
Here, the three parameters $\xi_{n}, \gamma_{n}$ and $ \theta_{n}$
are determined by the Bogoliubov coefficients as
\begin{eqnarray}
\xi_{n} = - i\frac{\nu^*_{n}}{\mu_{n}}, \quad \gamma_{n} = -2\ln(|\mu_n|), \quad e^{2 i\theta_{n}} = \frac{\mu_n^*}{\mu_n}. \label{spinor-parameters}
\end{eqnarray}
Note that the pair production on spinor QED is restricted to only
one pair of particle and antiparticle for a given quantum number $n$ due to the Pauli exclusion principle.
Thus, the mean number of pairs created
from the vacuum for each state $n$ at zero temperature is
calculated as
\begin{eqnarray}
\overline{\cal N}_{n}^{\rm sp}(T=0) = | \langle 1_{n},{\rm out} \vert 0,{\rm in} \rangle |^{2} = |\nu_{n}|^{2}.
\label{sp-mean}
\end{eqnarray}

\section{Pair Production at $T \neq 0$}

We now calculate the mean
number of pairs at finite temperature from the probability for each transition, as in the case of zero temperature. As there is no
mode-mixing, we separately calculate the mean number of created
pairs for each mode. For an initial thermal ensemble at $\beta =
1/kT$, which might not be charge neutral, the mean number of
produced pairs consists of the vacuum pair production, the induced
pair production, and the stimulated pair annihilation as shown in Fig. 1:
\begin{eqnarray}
\overline{\cal N}_{n} (T) &=& \frac{1}{Z_{n}} \Biggl[\sum_{n_{n} >
0}^{\infty} e^{-\beta E_{0_{n},0_{n}}} n_{n} P_{0_{n},0_{n}
\rightarrow n_{n},n_{n}} - \sum_{p_{n}
> n_{n} \ge 0, q_{n}
> m_{n} \ge 0}^{\infty} e^{-\beta E_{p_{n},q_{n}}} (p_{n} - n_{n})
P_{p_{n},q_{n} \rightarrow n_{n},m_{n}} \nonumber \\ &~&~~~~~~~~ +
\sum_{p_{n} > n_{n} \ge 0, q_{n}
> m_{n} \ge 0, n_{n} + m_{n} \neq 0}^{\infty} e^{-\beta E_{n_{n},m_{n}}} (p_{n}
- n_{n}) P_{n_{n},m_{n} \rightarrow p_{n},q_{n}} \Biggr],
\label{mean-T1}
\end{eqnarray}
where $Z_n$ is the partition function for the initial ensemble and
 $P_{n_{n},m_{n} \rightarrow p_{n},q_{n}} = |\langle
p_{n},q_{n}, {\rm in} \vert U_{n} \vert n_{n},m_{n}, {\rm in}
\rangle|^2$ is the transition probability from $\vert n_{n},m_{n},
{\rm in} \rangle$ to $\vert p_{n}, q_{n},{\rm in} \rangle$.
Using
$|\langle p_{n},q_{n}, {\rm in} \vert U_{n} \vert n_{n},m_{n},
{\rm in} \rangle|^2 = |\langle p_{n},q_{n}, {\rm in} \vert
U_{n}^{\dagger} \vert n_{n},m_{n}, {\rm in} \rangle|^2$, which
implies $P_{p_{n},q_{n} \rightarrow n_{n},m_{n}} = P_{n_{n},m_{n}
\rightarrow p_{n},q_{n}}$, the mean number of pairs created at
finite temperature, Eq. (\ref{mean-T1}),  can be written as
\begin{eqnarray}
\overline{\cal N}_{n} (T) &=& \frac{1}{2Z_{n}} \sum_{p_{n} >
n_{n},q_{n} > m_{n}}^{\infty} \Biggl[ \langle n_{n},m_{n}, {\rm
in} \vert e^{ - \beta H_{n,{\rm in}}} U_{n} \vert p_{n},q_{n},
{\rm in} \rangle \langle p_{n},q_{n}, {\rm in}\vert (N_{n,{\rm
in}}U_{n}^{\dagger}-U_{n}^{\dagger} N_{n,{\rm in}}) \vert
n_{n},m_{n}, {\rm in} \rangle \nonumber \\ &~&~~~~~~~~ - \langle
n_{n},m_{n}, {\rm in} \vert U_{n}^{\dagger} e^{-\beta H_{n,{\rm
in}}} \vert p_{n},q_{n}, {\rm in} \rangle \langle p_{n},q_{n},
{\rm in}\vert (N_{n,{\rm in}}U_{n}-U_{n} N_{n,{\rm in}}) \vert
n_{n},m_{n}, {\rm in} \rangle \Biggr] \nonumber \\ ~ &=&
\frac{1}{2Z_{n}} {\rm Tr}(U_{n,{\rm in}} N_{n,{\rm in}}U_{n,{\rm
in}}^{\dagger} -N_{n,{\rm in}}) e^{-\beta H_{n,{\rm in}}},
\label{mean-T2}
\end{eqnarray}
where $N_{n, {\rm in}} = a_{{\bf k}, {\rm in}}^{\dagger} a_{{\bf k},
{\rm in}} + b_{{\bf k}, {\rm in}}^{\dagger} b_{{\bf k}, {\rm in}}$
for scalar particles and $N_{n, {\rm in}} = b_{n, {\rm
in}}^{\dagger} b_{n, {\rm in}} + d_{n, {\rm in}}^{\dagger} d_{n,
{\rm in}}$ for spinor particles and $H_{n,{\rm in}}=\omega_{n,{\rm in}} N_{n, {\rm in}}$ for both.

\begin{figure}[htp]
\begin{center}
\includegraphics[width=5.2in]{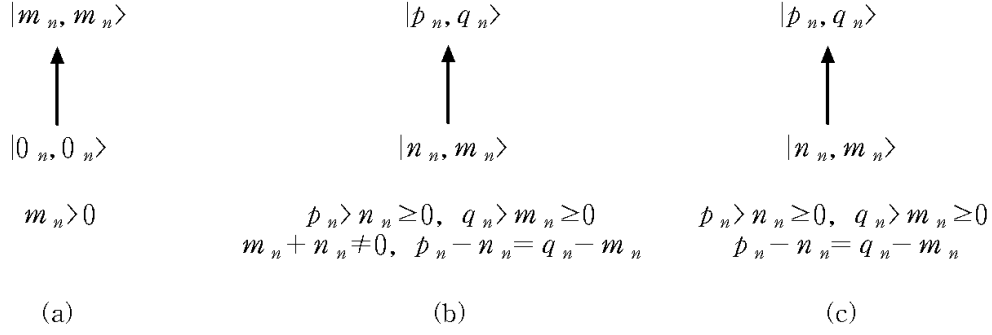}
\caption{(a) the vacuum production, (b) the induced production, and
(c) the stimulated annihilation}
\end{center}
\end{figure}

From the Bogoliubov transformations (\ref{sc-U-tr}) and (\ref{sp-U-tr}), we have $N_{n, {\rm out}}
= U_{n} N_{n, {\rm in}} U_{n}^{\dagger}$. Thus, the mean number of
pairs created at finite temperature can
be written concisely as
\begin{eqnarray}
\overline{\cal N}_{n}(T) &=& \frac{1}{2Z_{n}} {\rm Tr}(N_{n,{\rm
out}} - N_{n,{\rm in}}) e^{-\beta H_{n,{\rm in}}}. \label{mean-T3}
\end{eqnarray}
First, for scalar particles, from Eq. (\ref{sc-out-in}),
\begin{eqnarray}
N_{{\bf k},{\rm out}}-N_{{\bf k},{\rm in}}= 2|\nu_{\bf
k}|^{2}(a_{{\bf k},{\rm in}}^{\dagger}a_{{\bf k},{\rm in}}+b_{{\bf
k},{\rm in}}^{\dagger}b_{{\bf k},{\rm in}}+1)+2\mu_{\bf k}\nu_{\bf
k} a_{{\bf k},{\rm in}} b_{{\bf k},{\rm in}}+2\mu_{\bf
k}^{*}\nu_{\bf k}^{*} a_{{\bf k},{\rm in}}^{\dagger} b_{{\bf k},{\rm
in}}^{\dagger},
\end{eqnarray}
and $Z_{\bf k}= e^{\beta\omega_{{\bf k},{\rm in}}}(2\sinh
\beta\omega_{{\bf k},{\rm in}}/2)^{-2}$, the mean number of scalar pairs
created in momentum ${\bf k}$ at finite temperature is given by
\begin{eqnarray}
\overline{\cal N}_{\bf k}^{\rm sc}(T)=\frac{\vert\nu_{\bf
k}\vert^{2}}{Z_{\bf k}}\sum_{n,m=0}^{\infty}\langle
n,m\vert(n+m+1)e^{-(n+m)\beta\omega_{{\bf k},{\rm in}}}|n,m
\rangle= \vert \nu_{\bf k}\vert^{2}\coth\frac{\beta\omega_{{\bf
k},{\rm in}}}{2}. \label{sc-result}
\end{eqnarray}

Second, for spinor particles, from Eq. (\ref{sp-out-in}),
\begin{eqnarray}
N_{n,{\rm out}}-N_{n,{\rm in}} = -2|\nu_{n}|^{2}(b_{n,{\rm
in}}^{\dagger}b_{n,{\rm in}}+d_{n,{\rm in}}^{\dagger}d_{n,{\rm
in}}-1)+2i\mu_{n}\nu_{n} b_{n,{\rm in}} d_{n,{\rm
in}}+2i\mu_{n}^{*}\nu_{n}^{*} b_{n,{\rm in}}^{\dagger} d_{n,{\rm
in}}^{\dagger},
\end{eqnarray}
and $Z_{n}= (1+e^{-\beta\omega_{n,{\rm in}}})^{2}$, we find the mean
number of spinor pairs created in state $n$ at finite temperature
\begin{eqnarray}
\overline{\cal N}_{n}^{\rm
sp}(T)=-\frac{|\nu_{n}|^{2}}{Z_{n}}\sum_{n,m=0}^{1} \langle n,m
\vert (n+m-1)e^{-(n+m)
\beta\omega_{n,{\rm in}}} \vert n,m \rangle = \vert\nu_{n}\vert^{2} \tanh\frac{\beta\omega_{n,{\rm in}}}{2}.
\label{sp-result}
\end{eqnarray}

\begin{figure}[htp]
\begin{center}
\includegraphics[width=2.8in]{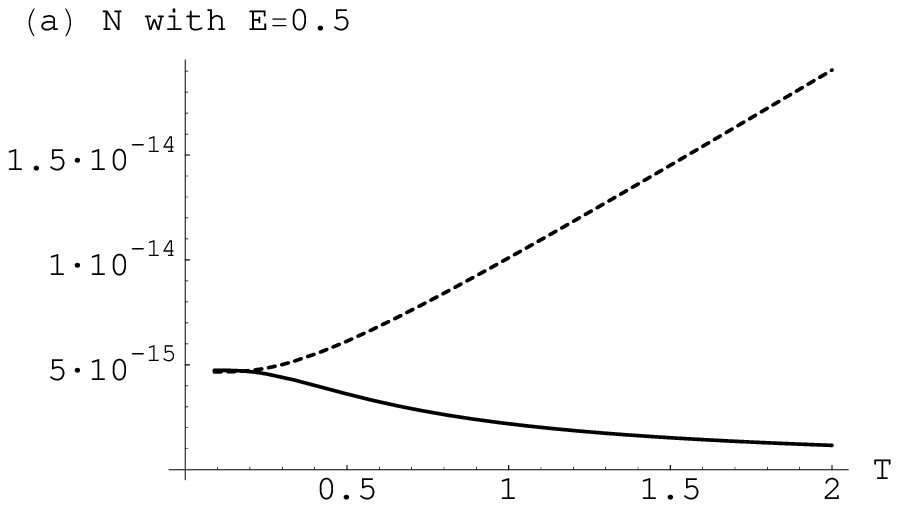} \hspace{3mm}
\includegraphics[width=2.8in]{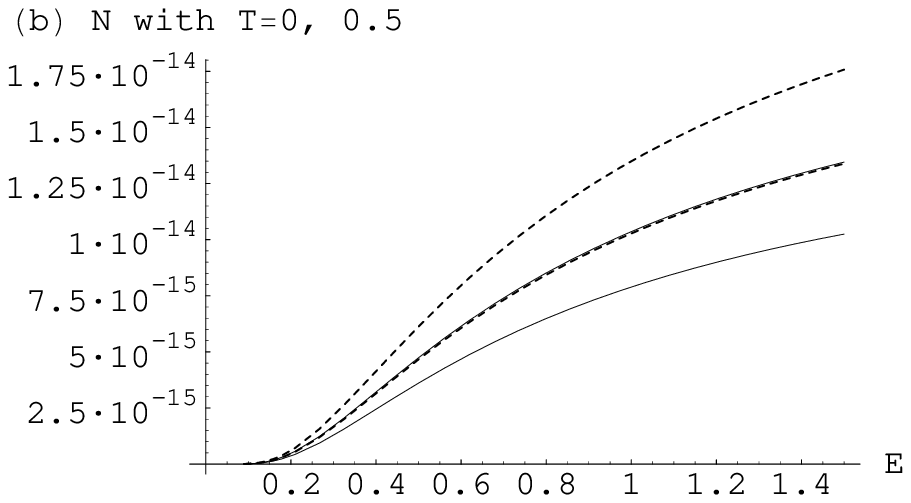} \\
\caption{The mean number of scalar(dashed line) and spinor(solid line) pairs created, $N$, in ${\bf k}=0$ state for a
pulsed Sauter-type electric field of duration t$=10$: (a) dependence on temperature $T$ with E$=0.5$, and (b)
dependence on electric field strength $E$ with T$=0.5$ for the upper and the lower lines, and with T$=0$ for the middle
lines, where dimensionless parameters $k_{B}T/mc^{2} \rightarrow $T, $qE\hbar/m^{2}c^{3} \rightarrow $E, and $m \tau
c^2/\hbar \rightarrow $t are used.}
\end{center}
\end{figure}

As an interesting model for discussions, we consider the Sauter-type electric field ${\cal E}(t) = E {\rm sech}^2
(t/\tau)$ with the gauge choice, $A_z (t) = - E \tau (1+ \tanh (t/\tau))$, which allows the exact solution leading to
the exact mean number of pairs created, $|\nu_{\bf k}|^{2}$, at zero temperature both in scalar and spinor QED
\cite{NN,GG,KLY,Kim-Lee07}. Also the approximation scheme of the WKB or worldline instanton method has been developed
\cite{Kim-Page,DSWG,KRX}. Because the Sauter-type electric field acts effectively for a finite period of time $\tau$,
the ingoing thermal states are stable and well defined as required in our formalism. Fig. 2 shows the mean number of
created pairs Eqs.(\ref{sc-result},\ref{sp-result}) at finite temperature with the electric field $E$, the duration
$\tau$ and temperature $k_B T = 1/\beta$ scaled in terms of the critical strength, the Compton time and the electron
mass, respectively. The production of scalar pairs is thermally enhanced, while the production of fermion pairs is
thermally suppressed as expected by the Pauli blocking, which is consistent with the calculation by density matrix
method \cite{GGT}.

The pure thermal effect on the mean number of created pairs, which is $\Delta \overline{\cal N}_{\bf k}(T) =
\overline{\cal N}_{\bf k}(T) - \overline{\cal N}_{\bf k}(0)$, is given by $\Delta \overline{\cal N}_{\bf k}(T) = \pm 2
|\nu_{\bf k}|^2 f^\frac{B}{F}_{\bf k}(T)$ with $f^\frac{B}{F}_{\bf k}(T)$ being the Bose-Einstein or Fermi-Dirac
distribution. For $k_B T \ll \omega_{{\bf k}, {\rm in}}$, the distribution $f^\frac{B}{F}_{\bf k}(T)$ approximately
equals to the Boltzmann factor $f_{\bf k} \approx e^{-\sqrt{m^{2}+{\bf k}^{2}} /k_B T}$. Thus, the mean number of
created pairs at finite temperature, Eqs.(\ref{sc-result},\ref{sp-result}), is reduced to the zero temperature result,
Eqs.(\ref{sc-mean},\ref{sp-mean}), for a temperature much lower than the rest mass.

Our result Eq.(\ref{sp-result}) could be compared with the calculation in imaginary-time formalism \cite{Gies99}, where
the thermal effect appears only at two-loop because thermal one-loop fluctuations are on-shell. On the other hand, Eq.
(\ref{sp-result}), an off-shell calculation, is equivalent to the thermal loop times the vacuum one-loop, in fact, part
of two-loops. One interesting comment to be pointed out is that the momentum integral over the distribution function in
$\Delta \overline{\cal N}_{\bf k}(T) = \pm 2 |\nu_{\bf k}|^2 f^\frac{B}{F}_{\bf k} (T)$ leads to the factor $T^4$ in
Ref. \cite{Gies99}, though the momentum integral of the distribution is intertwined with the vacuum pair-production
rate. To show rigorously the connection between our result and Ref. \cite{Gies99} requires calculating the effective
action at finite temperature along the line of Ref.\cite{KLY}, which will be addressed in the future.

\section{Conclusion}

In this paper, using the evolution operator method, we found that the mean number of created pairs in state $n$ at finite temperature is given by
\begin{eqnarray}
\overline{\cal N}_{n} (T) &=& \frac{1}{2Z_{n}} {\rm Tr}(N_{n,{\rm
out}} - N_{n,{\rm in}}) e^{-\beta H_{n,{\rm in}}} .
\end{eqnarray}
For scalar and spinor QED in external electric fields, the total mean number density of created pairs at finite temperature is given by
\begin{eqnarray}
\overline{\cal N}^{\rm sc} (T) &=& \int \frac{d^{3} {\bf
k}}{(2\pi)^{3}} \overline{\cal N}^{\rm sc}_{\bf k} (0)
\coth\frac{\beta\omega_{{\bf k},{\rm in}}}{2} \quad ({\rm for~
scalar}), \\ \overline{\cal N}^{\rm sp} (T) &=& \sum_{\sigma}\int
\frac{d^{3} {\bf k}}{(2\pi)^{3}}\overline{\cal N}^{\rm sp}_{n} (0)
\tanh\frac{\beta\omega_{n,{\rm in}}}{2} \quad ({\rm for~ spinor}),
\end{eqnarray}
where $\overline{\cal N}^{\rm sc}_{\bf k} (0)$ and $\overline{\cal
N}^{\rm sp}_{n} (0)$ are the mean number of created pairs at zero
temperature for scalar and spinor QED, respectively.

\acknowledgments

S.~P.~K. would like to appreciate the hospitality of Hanyang
University, and H.~K.~L. and Y.~Y. would like to appreciate the
hospitality of Kunsan National University. The work of S.~P.~K.
was supported by the Korea Research Foundation Grant funded by the
Korean Government (MOEHRD) (KRF-2007-C00167) and the work of
H.~K.~L. was supported by the Korea Science and Engineering
Foundation (KOSEF) grant funded by the Korea government (MOST)
(No. R01-2006-000-10651-0).

\appendix

\end{document}